# Long-range Scanning Tunneling Microscope for the study of nanostructures on insulating substrates

Aday Molina-Mendoza[1,a)], José Gabriel Rodrigo[1,2], Joshua Island[3], Enrique Burzurí[3], Gabino Rubio-Bollinger[1,2], Herre S. J. van der Zant[3], and Nicolás Agraït[1,2,4].

[1]*Departamento de Física de la Materia Condensada, Universidad Autónoma de Madrid, Campus de Cantoblanco, E-28049 Madrid, Spain*
[2]*Condensed Matter Physics Center (IFIMAC) and Instituto Universitario de Ciencia de Materiales "Nicolás Cabrera", Universidad Autónoma de Madrid, Campus de Cantoblanco, E-28049 Madrid, Spain*
[3]*Kavli Institute of Nanoscience, Delft University of Technology, P.O. Box 5046 2600 GA Delft, The Netherlands*
[4]*Instituto Madrileño de Estudios Avanzados en Nanociencia IMDEA-Nanociencia, E-28049 Madrid, Spain*

The Scanning Tunneling Microscope is a powerful tool for studying the electronic properties at the atomic level, however it's relatively small scanning range and the fact that it can only operate on conducting samples prevents its application to study heterogeneous samples consisting on conducting and insulating regions. Here we present a long-range scanning tunneling microscope capable of detecting conducting micro and nanostructures on insulating substrates using a technique based on the capacitance between the tip and the sample and performing STM studies.

## I. INTRODUCTION

The study at the atomic level of the electronic properties and electrical transport of heterogeneous surfaces such as nanopatterned structures or atomically thin 2D-crystals like graphene, $MoS_2$ or $TaSe_2$, requires placing a local probe, i.e. STM tip, on small conducting regions (typically a few micrometers in size) surrounded by an insulating substrate. In principle, locating the conductive regions can be achieved using a long-range optical microscope[1], a scanning electron microscope[2] or Raman spectroscopy[3,4], however these approaches have the important drawback of being difficult to implement in a low temperature scanning tunneling microscope. A solution to this problem is the recently reported method by Li *et al.*[5] based on the measurement of the capacitance between the tip and the sample, making it possible to navigate safely to a conductive region on an insulating surface.

Here, we present a scanning tunneling microscope compatible with low temperature operation implementing the aforementioned capacitance method. Our design features a precise *xy* motion in a range of up to 5 mm, and is able to perform fast long-range non-contact capacitive scans. A conventional piezotube is used for STM mode scans in a range of 1μm. We demonstrate these capabilities locating in capacitance mode and imaging in STM mode graphene flakes connected to two different electrodes.

## II. INSTRUMENT DESIGN

The STM head consists of two separate parts, each one containing a linear positioning stage. The lower part houses the *x-axis* positioner and the upper the *y-axis* positioner. Each of these linear positioners consists of four multilayer shear piezostacks actuators (PI Piezotechnology) placed in a V-shaped channel constraining the motion of a slider to one dimension. These linear positioners are similar to the design reported by Pan *et al.*[6] with the important difference that instead of having piezostacks on all three sides, the slider is held into position by a spring, leaving one of its lateral surfaces free. These free surfaces are used to attach a piezotube scanner, holding the tip, to the *y-axis* positioner (upper part) and a sample-holder to the *x-axis* positioner (lower part) (see Fig. 1).

The *x-axis* and *y-axis* positioners can operate either in a purely inertial mode, in which the same ramp is applied to all 4 stacks or in a mixed inertial-step fashion in which the ramp is slightly delayed at each piezostack using resistors (Fig. 2). Step sizes range from a few nanometers to 300 nm. The total displacement range is up to 5 mm.

Coarse approach in the vertical direction (*z-axis*) is achieved using three micrometer screws placed in the upper part. Two of them are used for the initial tip-sample coarse positioning and the third one is powered by a piezoelectric ultrasonic motor[7], enabling automated tip-sample approach to operating conditions (see Fig. 1).

---

a) aday.molina@uam.es

Upper and lower parts are held together by means of springs (not shown in Fig. 1).

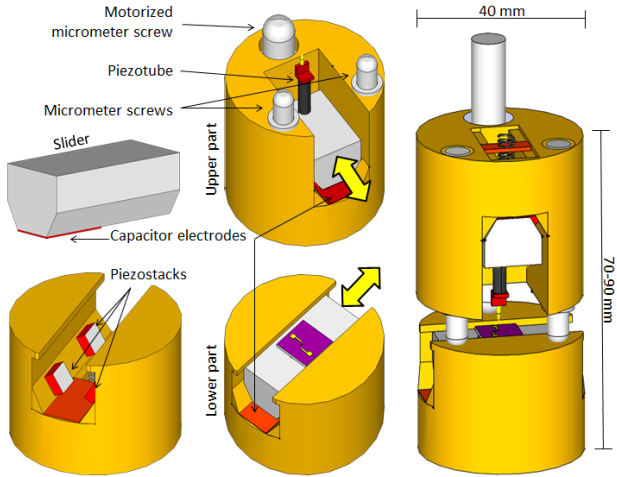

FIG. 1 Drawing of the different parts of the microscope and the assembled STM head. Left column: one of the sliders and the lower part of the microscope, showing the piezostacks. Central column: the two main parts of the STM head, each one with the corresponding slider. Right column: a drawing of the assembled STM head.

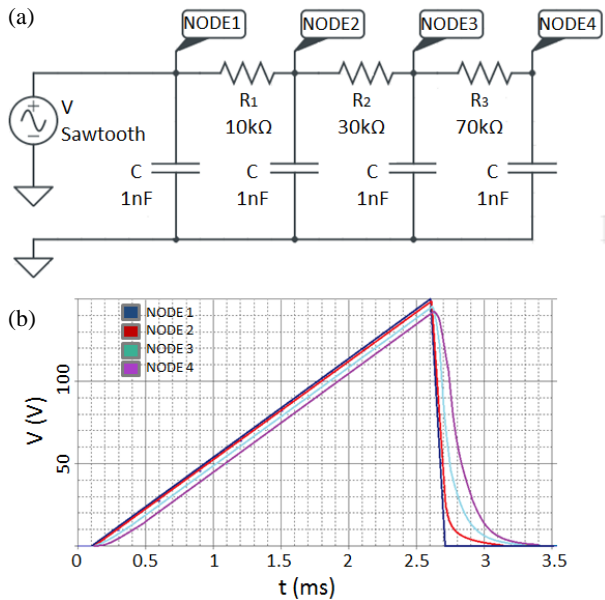

FIG. 2 (a) Schematic of the equivalent electric circuit for the piezoelectric positioners. Each piezostack is represented here by a capacitor of C = 1nF. The voltage signal applied is sawtooth-like. (b) Graph representing the voltage signal at each piezostack. As it can be seen, the signal at each node is delayed with respect to the others.

The displacement of the *x* and *y* sliders is obtained by measuring the capacitance of two overlapping metal plates, one of which is attached to the slider and the other one to the body of the microscope. Using a capacitance bridge (Andeen Hagerling 2500A) it is possible to determine the absolute position of the slider within 1 μm in the whole range of motion (5 mm).

This STM has been designed having in mind its implementation in a cryogenic system, consequently all dimensions have been kept small enough to fit in a $^3$He refrigerator. All the components and positioning stages have been tested separately at temperatures in the range of the milikelvin.

## III. EXPERIMENT

We demonstrate the capabilities of our new STM design by exploring in STM mode graphene flakes in which a nanogap is opened by electroburning in ambient conditions[8]. The graphene flakes are deposited by mechanical exfoliation of kish graphite on degenerately doped silicon substrates coated with 280nm of thermal silicon dioxide, and electrically connected by Au electrodes patterned by electron-beam lithography and metal evaporation (Fig. 3).

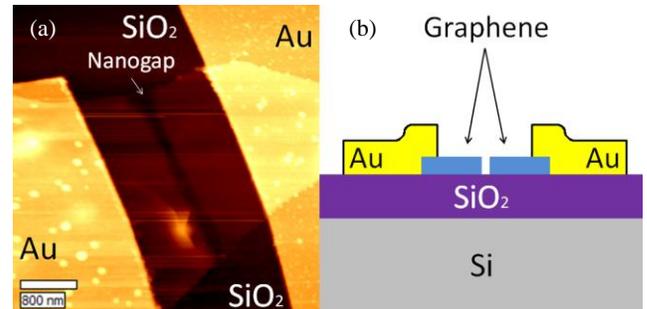

FIG. 3 (a) AFM image of a graphene flake presenting a nanogap opened by electroburning and the connecting gold electrodes. (b) Schematic of the sample structure.

The capacitance based technique introduced by *Li et. al.*[5] is used to locate the graphene flakes. Specifically, we perform long-range scans (hundreds of microns) using the *xy*-positioning stages and registering the capacitance between the tip and each of the two electrodes contacting the flake. Thus it is possible to follow the electrodes leading to the flake. Capacitive scans can be performed at heights lower than the 30% of the width of the conducting elements[5], avoiding the risk of crashing the tip onto the sample.

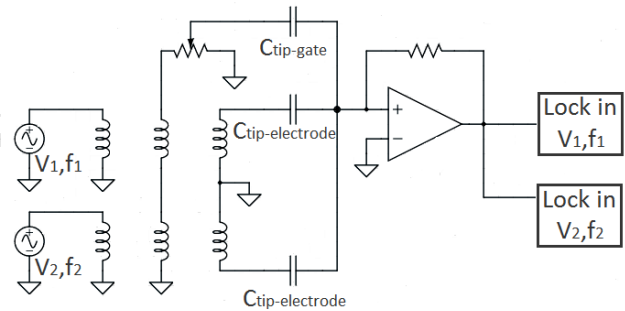

FIG. 4 Schematic of the circuit used for detecting the electrodes. The three capacitors represent the capacitances between the tip and the gate, and between the tip and each of the two electrodes.

We apply AC voltages ($V_1$ and $V_2$) to the Au electrodes with different frequency for each one, and the sum of them with opposite sign to the silicon substrate (*back-gate*). The capacitive current induced on the tip is measured with two lock-in amplifiers, each one tuned to one of the frequencies applied to the electrodes, allowing to obtain independently the capacitance between the tip and each of the two electrodes (Fig. 4). The signal between the tip and the *back-gate* can be adjusted with a potentiometer to enhance resolution.

In order to have higher lateral resolution, the tips used in this system (Fig. 5) consist of a thin carbon fibre (ø = 7μm, L = 120μm) glued to a Pt-Ir wire (ø = 250μm, L~3mm) with silver epoxy. The carbon fibres are electrochemically etched before being glued, obtaining a radius of curvature at the apex of the order of 55nm[9]. The use of such a thin tip enhances the contrast in capacitance across the different regions of the sample.

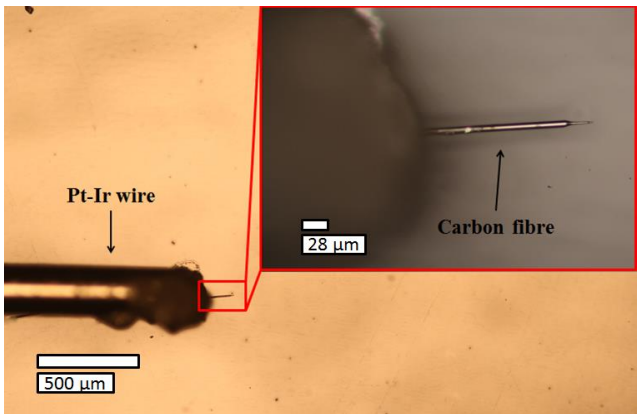

FIG. 5 Pictures taken with an optical microscope of a carbon fibre tip electrochemically etched. The Pt-Ir wire has a diameter of 250μm, while the carbon fibre has a diameter of 7μm.

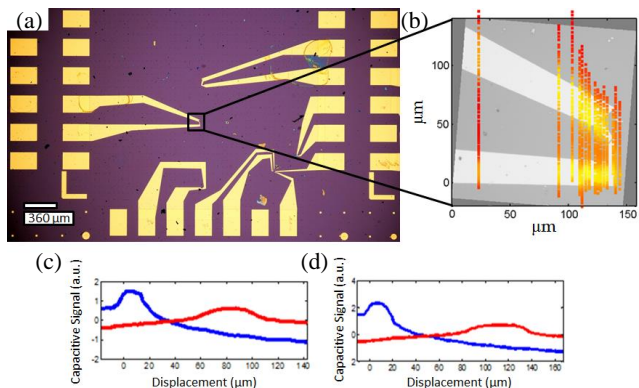

FIG. 6 (a) Optical microscope image of the sample. (b) Series of capacitive scans superposed to the optical image. The scan lines are obtained adding the output of the two lock-in amplifiers. The red/yellow dots correspond to lower/higher capacitance. (c) and (d) show the capacitance signal along two of the scans in (b), for each scan we present the output of the two lock-in amplifiers as a function of tip displacement. The blue/red curve corresponds to the capacitance between the tip and the lower/upper electrode.

In Fig. 6 we illustrate how the capacitive scans are used to identify the electrodes and locate the flake. Once the flake is located, we switch to the standard STM mode to obtain nanometer-scale resolution using the piezotube scanner (Fig.7).

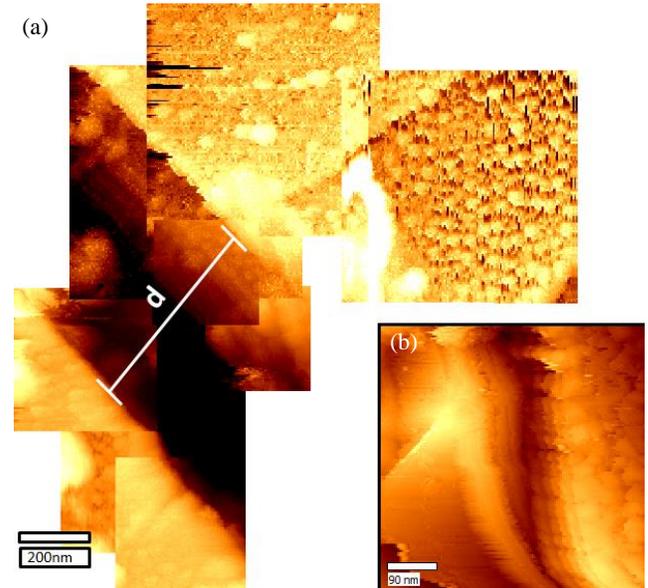

FIG 7 (a) Composition of STM images of a graphene flake with a nanogap in the middle between the two gold electrodes corresponding to the area shown in Fig. 6. The distance between the edges of the two Au electrodes is d = 590nm. (b) STM image of the nanogap in another flake where different layers can be seen. The measured depth of the gap is 11.5nm.

## V. SUMMARY

We have constructed a Scanning Tunneling Microscope which combines a long-range capacitive imaging mode with the standard STM imaging mode. The instrument is capable of performing scans in the capacitive mode (from 1μm up to few mm), detecting conducting surfaces on insulating substrates and performing standard STM studies in the selected conducting regions. It has been used for locating and investigating with STM graphene flakes on a silicon dioxide substrate.

## ACKNOWLEDGEMENTS

We want to thank Santiago Márquez for the construction of the microscope. This work was supported by MICINN/MINECO (Spain) through the programs MAT2011-25046 BES-2012-057346, FIS2011-23488 and CONSOLIDER-INGENIO-2010 CSD-2007-00010, Comunidad de Madrid (Spain) through the program NANOBIOMAGNET (s2009/MAT-1726), the Dutch organization for Fundamental Research of Matter (FOM), and by the EU through the network "ELFOS" (FP7-ICT2009-6) and the ERC Advanced grant (Mols@Mols).